\documentclass[useAMS,usenatbib]{mn2e}

\usepackage{amsmath}
\usepackage{amssymb}
\usepackage{graphicx}
\usepackage{url}

\usepackage[pass]{geometry}

\newcommand{\xa}{\ensuremath{x^{\rm a}}}
\newcommand{\xb}{\ensuremath{x^{\rm b}}}

\newcommand{\xk}{\ensuremath{x^{k}}}
\newcommand{\xkp}{\ensuremath{x^{k+1}}}
\newcommand{\xkm}{\ensuremath{x^{k-1}}}

\newcommand{\na}{\ensuremath{n_{\rm a}}}
\newcommand{\nb}{\ensuremath{n_{\rm b}}}

\newcommand{\discrim}{\ensuremath{\mathcal{D}}}

\newcommand{\Dx}{\ensuremath{\Delta x}}
\newcommand{\Dxk}{\ensuremath{\Delta \xk}}

\newcommand{\wk}{\ensuremath{w_{k}}}

\newcommand{\lambdai}{\ensuremath{\lambda^{i}}}
\newcommand{\lambdaAi}{\ensuremath{\lambdai_{\mathsf{A}}}}

\newcommand{\Dnu}{\ensuremath{\Delta\nu_{2}}}
\newcommand{\dnu}{\ensuremath{\delta\nu_{13}}}

\newcommand{\omegaex}{\ensuremath{\omega_{\rm ex}}}

\newcommand{\omegaa}{\ensuremath{\omega_{\rm a}}}
\newcommand{\omegab}{\ensuremath{\omega_{\rm b}}}

\newcommand{\pone}{p$_{1}$}

\newcommand{\order}{\ensuremath{o}}

\newcommand{\error}{\ensuremath{\epsilon}}

\newcommand{\inertia}{\ensuremath{\mathcal{E}}}
\newcommand{\norminertia}{\ensuremath{E}}

\newcommand{\npr}{\ensuremath{n_{\rm p}}}
\newcommand{\ngr}{\ensuremath{n_{\rm g}}}

\newcommand{\ncore}{\ensuremath{n_{\rm core}}}
\newcommand{\nnode}{\ensuremath{n_{\rm node}}}
\newcommand{\nproc}{\ensuremath{n_{\rm proc}}}

\newcommand{\Dfreq}{\ensuremath{\Delta \nu}}

\newcommand{\dP}{\ensuremath{\Delta \Pi}}

\newcommand{\numax}{\ensuremath{\nu_{\rm max}}}

\newcommand{\tdyn}{\ensuremath{\tau}}

\newcommand{\vy}{\ensuremath{\mathbf{y}}}

\newcommand{\vyk}{\ensuremath{\vy^{k}}}

\newcommand{\vunk}{\ensuremath{\mathbf{u}}}
\newcommand{\vunkab}{\ensuremath{\mathbf{u}^{\rm c}}}
\newcommand{\vunkred}{\ensuremath{\tilde{\mathbf{u}}}}

\newcommand{\vnull}{\ensuremath{\mathbf{0}}}

\newcommand{\mjac}{\ensuremath{\mathbfss{A}}}
\newcommand{\mjacnorm}{\ensuremath{\tilde{\mjac}}}

\newcommand{\mfund}{\ensuremath{\mathbfss{Y}}}

\newcommand{\mbounda}{\ensuremath{\mathbfss{B}^{\rm a}}}
\newcommand{\mboundb}{\ensuremath{\mathbfss{B}^{\rm b}}}

\newcommand{\msys}{\ensuremath{\mathbfss{S}}}
\newcommand{\msyscompact}{\ensuremath{\mathbfss{S}^{\rm c}}}

\newcommand{\mident}{\ensuremath{\mathbfss{I}}}
\newcommand{\mnull}{\ensuremath{\mathbfss{0}}}

\newcommand{\mmagnus}{\ensuremath{\mathbf{\Omega}}}
\newcommand{\malpha}{\ensuremath{\balpha}}
\newcommand{\mC}{\ensuremath{\mathbfss{C}}}

\newcommand{\meigvalA}{\ensuremath{\mathbf{\Lambda}_{\mathsf{A}}}}
\newcommand{\meigvalO}{\ensuremath{\mathbf{\Lambda}_{\Omega}}}

\newcommand{\meigvecA}{\ensuremath{\mathbfss{M}_{\mathsf{A}}}}
\newcommand{\meigvecO}{\ensuremath{\mathbfss{M}_{\Omega}}}

\newcommand{\mU}{\ensuremath{\mathbfss{U}}}
\newcommand{\mUred}{\ensuremath{\tilde{\mU}}}
\newcommand{\mL}{\ensuremath{\mathbfss{L}}}

\newcommand{\diff}{\ensuremath{\mathrm{d}}}
\newcommand{\field}[1]{\ensuremath{\mathbb{#1}}}

\newcommand{\Msun}{\ensuremath{M_{\odot}}}

\newcommand{\gram}{\ensuremath{\mathrm{g}}}
\newcommand{\cm}{\ensuremath{\mathrm{cm}}}
\newcommand{\second}{\ensuremath{\mathrm{s}}}

\newcommand{\Gyr}{\ensuremath{\mathrm{Gyr}}}

\newcommand{\muHz}{\ensuremath{\mu\mathrm{Hz}}}
\newcommand{\nHz}{\ensuremath{\mathrm{nHz}}}

\newcommand{\gyre}{GYRE}

\newcommand{\mesastar}{MESA\,STAR}

\newcommand{\adipls}{ADIPLS}
\newcommand{\astec}{ASTEC}
\newcommand{\boojum}{BOOJUM}
\newcommand{\graco}{GraCo}
\newcommand{\posc}{POSC}
\newcommand{\noc}{NOC}
\newcommand{\lnawenr}{LNAWENR}
\newcommand{\oscrox}{OSCROX}
\newcommand{\osc}{OSC}
\newcommand{\mad}{MAD}
\newcommand{\filou}{FILOU}
\newcommand{\pulse}{PULSE}

\newcommand{\lapack}{LAPACK}

\newcommand{\blas}{BLAS}

\newcommand{\kth}{\ensuremath{k^{\rm th}}}

\newcommand{\BV}{Brunt-V\"ais\"al\"a}

\newcommand{\most}{\textit{MOST}}
\newcommand{\corot}{\textit{CoRoT}}
\newcommand{\kepler}{\textit{Kepler}}

\title[GYRE: An open-source stellar oscillation code]{GYRE: An open-source stellar oscillation code based on a new Magnus Multiple Shooting Scheme}
\author[R. H. D. Townsend and S. A. Teitler]
{R. H. D. Townsend\thanks{E-mail: townsend@astro.wisc.edu} and
 S. A. Teitler\\
Department of Astronomy, University of Wisconsin-Madison, 
2535 Sterling Hall, 475 N. Charter Street, Madison, WI 53706, USA}

\begin{document}

\date{Accepted \qquad Received}

\pagerange{\pageref{firstpage}--\pageref{lastpage}} \pubyear{2013}

\maketitle

\label{firstpage}

\begin{abstract}

We present a new oscillation code, \gyre, which solves the stellar
pulsation equations (both adiabatic and non-adiabatic) using a novel
Magnus Multiple Shooting numerical scheme devised to overcome certain
weaknesses of the usual relaxation and shooting schemes appearing in
the literature.  The code is accurate (up to 6th order in the number
of grid points), robust, efficiently makes use of multiple processor
cores and/or nodes, and is freely available in source form for use and
distribution. We verify the code against analytic solutions and
results from other oscillation codes, in all cases finding good
agreement. Then, we use the code to explore how the asteroseismic
observables of a 1.5\,\Msun\ star change as it evolves through the
red-giant bump.
\end{abstract}


%% Sections

\section{Introduction} \label{sec:intro}

The field of asteroseismology has been reinvigorated in recent years
thanks to the wealth of new observational data provided by space-based
instruments. Over the past decade there have been three satellite
missions with specific asteroseismic objectives: \most\
\citep{Wal2003,Mat2007}, launched in 2003; \corot\
\citep{Mic2008,Bag2009}, launched in 2006; and \kepler\
\citep{Bor2009,Gil2010,Kje2010}, launched in 2009. Exciting results
from these missions include the discovery that nearly all
$\gamma$~Doradus and $\delta$~Scuti stars are hybrid pulsators
\citep{Gri2010}; ensemble asteroseismic analysis of solar-like
oscillations in hundreds of solar-type stars \citep{Cha2011}; and the
detection of solar-like oscillations in a large sample of red giants
\citep{DeR2009}. See also the reviews by \citet{Chr2011} and
\citet{ChaMig2013} for further highlights.

Interpreting these new observations requires the seismologist's analog
to the telescope: a stellar oscillation code which calculates the
eigenfrequency spectrum of an arbitrary input stellar model. Comparing
a calculated spectrum against a measured one provides a concrete
metric for evaluating a model, and therefore constitutes the bread and
butter of quantitative asteroseismology. Although the task of
iteratively improving model parameters has in the past been quite
cumbersome, there are now tools available that largely automate this
process. The Asteroseismic Modeling Portal \citep[AMP;][]{Met2009}
provides a web-based front end for asteroseismic analysis and model
optimization of solar-like stars, using the Aarhus Stellar Evolution
code \citep[\astec;][]{Chr2008a} to build models and the Aarhus
Pulsation code \citep[\adipls;][]{Chr2008b} to calculate their
eigenfrequencies. Likewise, the widely adopted MESA stellar evolution
code \citep{Pax2011,Pax2013} includes an asteroseismology module also
based on \adipls, and offering similar optimization capabilities to
AMP.

Such tools place ever-increasing demands on the oscillation codes that
underpin them. A code will typically be executed hundreds or thousands
of times during an optimization run, and must therefore make efficient
use of available computational resources (e.g., multiple processor
cores and/or cluster nodes). The code must be robust, running and
producing sensible output without manual intervention like
hand-tuning. The code must have an accuracy that matches or exceeds
the frequency precision now achievable by satellite missions. Finally,
it is preferable that the code address the various physical processes
that inevitably complicate calculations, such as non-adiabaticity,
rotation, and magnetic fields.

These desiderata motivated us to develop a new oscillation code,
\gyre, which we describe in the present paper. The code is based on a
`Magnus Multiple Shooting' (MMS) scheme for solving the linearized
pulsation equations, devised by us to address various pitfalls
encountered with the standard relaxation and shooting schemes
appearing in the literature. The following section reviews these
schemes and the existing oscillation codes which use
them. Section~\ref{sec:scheme} then describes the MMS scheme in
detail, and Section~\ref{sec:implement} discusses how the scheme is
implemented in \gyre. We present example calculations in
Section~\ref{sec:calc}, and discuss and summarize the paper in
Section~\ref{sec:discuss}.

%% Background

\section{Background} \label{sec:background}

The differential equations and algebraic boundary conditions governing
small-amplitude non-radial oscillations of a star about an equilibrium
background state --- the so-called linearized stellar pulsation
equations, presented in the adiabatic case in
Appendix~\ref{app:pulseqs} --- constitute a two-point boundary value
problem (BVP) in which the oscillation frequency $\omega$ serves as an
eigenvalue \citep[for a comprehensive review, see the monographs
  by][]{Cox1980,Unn1989,Aer2010,Sme2010}. Although there exist special
cases where analytic solutions exist \citep[e.g.,][]{Pek1938}, in
general this BVP must be solved numerically. Oscillation codes
specializing in this task were first described a half century ago, and
since then many different numerical schemes have been proposed in the
literature. The following sections review the two most prevalent, and
Section~\ref{ssec:background-other} then briefly discusses other
approaches which have been adopted.

\subsection{Relaxation Schemes} \label{ssec:background-relax}

Relaxation schemes for BVPs replace the derivatives in the
differential equations with finite-difference approximations specified
on a grid. Applied to pulsation problems, the finite-difference
relations together with boundary conditions and a normalization
condition establish a (typically large) system of algebraic equations
in which the unknowns are the dependent variables \vy\ at the grid
points plus the dimensionless oscillation frequency $\omega$. Because
these equations are non-linear (specifically, bi-linear in \vy\ and
$\omega^{2}$ for adiabatic pulsation), the simultaneous determination
of all unknowns requires iterative improvement of a trial solution ---
for instance using the procedure developed initially for stellar
evolution calculations by \citet{Hen1964}, which can be regarded as a
multi-dimensional Newton-Raphson algorithm. (\citealt{Unn1989} present
a detailed implementation of this procedure specifically tailored to
the pulsation equations).

The convergence of the Henyey scheme depends on how close a trial
solution is to a true solution. \citet{Cas1971} proposed an elegant
approach to finding good trial solutions for radial pulsation
problems, which \citet{OsaHan1973} subsequently adapted to the
non-radial case. One of the boundary conditions is set aside, allowing
the system of algebraic equations to be solved at any $\omega$ using a
standard \emph{linear} algorithm (e.g., Gaussian elimination). The
overlooked boundary condition is then used to construct a discriminant
function $\discrim(\omega)$ which vanishes when the boundary condition
is satisfied. Clearly, the roots of $\discrim(\omega)$ correspond to
the eigenfrequencies of the full BVP; thus, good trial solutions can
be obtained by isolating and refining these roots.

Relaxation using the \citet{Cas1971} approach has proven very popular,
forming the basis for many oscillation codes including the
\boojum\ code \citep{Tow2005}, the Nice Oscillation code
\citep[\noc;][]{Pro2008}, the Granada Oscillation code
\citep[\graco;][]{MoyGar2008}, and the \lnawenr\ code
\citep{Sur2008}. 
It is generally robust, but can run into difficulty when the
discriminant function exhibits singularities. These arise when the
dependent variable used for normalization naturally exhibits a zero at
the point where the normalization is applied. \citet{Unn1989} propose
addressing this problem by dividing the discriminant by one of the
dependent variables evaluated at the opposite boundary to the
overlooked boundary condition. This approach works well for the
adiabatic pulsation equations within the \citet{Cow1941} approximation
(where perturbations to the gravitational potential are neglected),
because neither of the dependent variables in this second-order BVP is
ever zero at the boundaries. However, in more general cases no such
guarantees can be made, and the division itself can make the
singularities recrudesce.

Attempts at more-sophisticated fixes to the singular discriminant
problem seem similarly doomed to failure. Because the singularities
ultimately stem from the imposition of an inappropriate normalization,
it is better to avoid normalization altogether when searching for
eigenfrequencies; this is the approach taken by the MMS scheme
(Section~\ref{sec:scheme}).

\subsection{Shooting Schemes} \label{ssec:background-shoot}

Shooting schemes treat BVPs as a set of initial value problems (IVPs),
with matching conditions applied where pairs of these IVPs meet. In the
stellar oscillation literature `double shooting' (also termed
`shooting to a fitting point') is most commonly encountered: IVPs are
integrated from each boundary toward an internal fitting point, with
initial values determined from the boundary conditions. The mismatch
between solutions at the fitting point is quantified by a discriminant
function $\discrim(\omega)$ which vanishes when the integrations
match. As before, the roots of $\discrim(\omega)$ correspond to the
eigenfrequencies of the BVP.

\citet{Hur1966} and \citet{Sme1966,Sme1967} were among the first to
apply double shooting to the pulsation equations. \citet{Scu1974}
adopted a simplified version of the scheme, where the fitting point is
placed at a boundary and only one IVP integration is performed
(so-called `single shooting' or `simple shooting'); however, the
integration can become unstable when approaching a boundary where the
differential equations become singular (i.e., the inner boundary, and
in polytropic models the outer boundary too). Modern oscillation codes
based on double shooting include \adipls\ (which can use either shooting
or relaxation), the Porto Oscillation Code \citep[\posc;][]{Mon2008}
and the \oscrox\ code \citep{Rox2008}.

\citet{Chr1980} discusses a complication that arises when using double
shooting to solve the adiabatic pulsation equations without the
\citet{Cow1941} approximation. This fourth-order BVP requires
integrating two linearly independent solutions from each boundary. In
evanescent regions these solutions are dominated by an exponentially
growing component, and they can easily become numerically linearly
dependent. The problem cannot be fixed by switching the direction of
integration (as one might do with an IVP), because the BVP has an
`exponential dichotomy' --- components that grow and decay
exponentially in both directions\footnote{In fact this is a good
  thing; as \citet{deHMat1987} demonstrate, an exponential dichotomy
  is a necessary condition for a BVP to be well conditioned.}. This is
a well-established weakness of single/double shooting schemes in
general, and has been extensively analyzed in the BVP literature
\citep[see, e.g., the excellent monograph by][]{Asc1995}. Happily, the
same literature provides a number of strategies for avoiding this
weakness. One of them, the Ricatti method, has already been used by
\citet{GauGla1990} to explore highly non-adiabatic oscillations
\citep[and see also][]{Val2013}. Another, multiple shooting, forms the
basis of the MMS scheme.

\subsection{Other Approaches} \label{ssec:background-other}

Although shooting and relaxation dominate in the stellar oscillation
literature, they are by no means the only schemes used. The
groundbreaking paper by \citet{Hur1966}, already mentioned above as an
early instance of shooting, also describes a collocation method (and
the authors allude to the possibility of a third approach, which can
be recognized as relaxation!). Collocation methods approximate BVP
solutions as a superposition of basis functions (e.g., Chebyshev
polynomials) which satisfy the differential equations exactly at a set
of nodes. Recently, \citet{Ree2006} again used collocation to explore
oscillations of polytropes, but this time incorporating the effects of
rapid rotation.

The finite element method (FEM) shares some similarities with
collocation methods, also using superpositions of basis
functions. However, the functions are chosen to minimize certain
integrals representative of the solution error. Two examples of
FEM-based oscillation codes are \filou\ \citep{SuaGou2008} and
\pulse\ \citep{BraCha2008}.

One other approach garnering some interest is inverse iteration. As
with relaxation, the differential equations are approximated with
finite differences. However, the resulting algebraic equations are
explicitly structured as a generalized linear eigenvalue problem,
which is then solved using the well-established technique of inverse
iteration \citep[e.g.,][]{GolVanL1996}. With a good trial solution
convergence is rapid. This approach is used by the \mad\ code
\citep{Dup2001} and the Li\`ege oscillation code
\citep[\osc;][]{Scu2008}.

%% Scheme

\section{The Magnus Multiple-Shooting Scheme} \label{sec:scheme}

\subsection{Problem Statement} \label{ssec:scheme-prob}

The MMS scheme solves BVPs defined by a system of linear, homogeneous,
first-order ordinary differential equations
\begin{equation} \label{eqn:diff-eqns}
\frac{\diff \vy}{\diff x} = \mjac(x) \vy
\end{equation}
defined on the interval $\xa \leq x \leq \xb$, together with boundary
conditions applied at each end of the interval,
\begin{equation} \label{eqn:bound-conds}
\mbounda \vy(\xa) = \vnull, \qquad \mboundb \vy(\xb) = \vnull.
\end{equation}
For $n$ equations, $\vy \in \field{C}^{n}$ is the vector of dependent
variables and $\mjac \in \field{C}^{n \times n}$ is the Jacobian
matrix. If \na\ of the boundary conditions are applied at the inner
point \xa\ and the remaining $\nb \equiv n-\na$ are applied at the
outer point \xb, then $\mbounda \in \field{C}^{\na \times n}$ and
$\mboundb \in \field{C}^{\nb \times n}$.

\subsection{Multiple Shooting} \label{ssec:scheme-multi}

Multiple shooting is a natural extension of the single/double shooting
schemes discussed in Section~\ref{ssec:background-shoot} which avoids
the numerical difficulties encountered when the system of equations
exhibits an exponential dichotomy. \citet{Asc1995} discuss it in
considerable depth; here, we highlight the important aspects.
The interval is divided up into a grid of $N$ points
\begin{equation} \label{eqn:x-grid}
\xa \equiv x^{1} < x^{2}< \ldots < x^{N-1}, x^{N} \equiv \xb.
\end{equation}
The solution to the BVP at any point in the \kth\ subinterval $\xk
\leq x \leq \xkp$ ($k = 1,2,\ldots,N-1$) is written as
\begin{equation} \label{eqn:ivp}
\vy(x) = \mfund(x;\xk) \vyk,
\end{equation}
where $\vyk \equiv \vy(\xk)$ and the fundamental solution
$\mfund(x;x') \in \field{C}^{n \times n}$ is the matrix function
satisfying the IVP
\begin{equation} \label{eqn:fund-sol}
\frac{\diff \mfund}{\diff x} = \mjac(x) \mfund, \qquad \mfund(x';x') = \mident.
\end{equation}
Here, $\mident$ is the rank-$n$ identity matrix.

The requirement that \vy\ be continuous at subinterval edges
imposes the matching condition
\begin{equation} \label{eqn:match}
\vy^{k+1} = \mfund^{k+1;k} \vyk,
\end{equation}
where we use the shorthand
\begin{equation}
\mfund^{k+1;k} \equiv \mfund(\xkp;\xk)
\end{equation}
for the fundamental solution matrix spanning the \kth\ subinterval.
There are $N-1$ such matching conditions, and in combination with the
boundary conditions~(\ref{eqn:bound-conds}) they lead to the system of
algebraic equations
\begin{equation} \label{eqn:alg-eqns}
\msys \, \vunk = \vnull.
\end{equation}
The vector of unknowns $\vunk \in \field{C}^{Nn}$ packs together the
dependent variables at the grid points,
\begin{equation} \label{eqn:vunk}
\vunk = 
\begin{pmatrix}
\vy^{1} \\
\vy^{2} \\
\vdots \\
\vy^{N-1} \\
\vy^{N}
\end{pmatrix},
\end{equation}
and the system matrix $\msys \in \field{C}^{Nn \times Nn}$ has a
staircase structure \citep[e.g.,][]{Fou1984} given by
\begin{equation} \label{eqn:smatrix}
\msys =
\begin{pmatrix}
\mbounda     & \mnull       & \mnull  & \cdots & \mnull         & \mnull \\
-\mfund^{2;1} & \mident      & \mnull  & \cdots  & \mnull         & \mnull \\
\mnull       & -\mfund^{3;2} & \mident & \cdots  & \mnull         & \mnull \\
\vdots       & \vdots       & \vdots  & \ddots & \vdots         & \vdots \\
\mnull       & \mnull       & \mnull  & \cdots & -\mfund^{N;N-1} & \mident \\
\mnull       & \mnull       & \mnull  & \cdots & \mnull         & \mboundb
\end{pmatrix}.
\end{equation}

As a linear homogeneous system, eqn.~(\ref{eqn:alg-eqns}) admits
non-trivial solutions only when the determinant of the system matrix
vanishes,
\begin{equation} \label{eqn:char}
\det (\msys) = 0.
\end{equation}
This can be recognized as the characteristic equation of the BVP. In
the case of the pulsation equations \msys\ depends implicitly on
$\omega$; thus, the stellar eigenfrequencies are the roots of the
discriminant function
\begin{equation} \label{eqn:discrim}
\discrim(\omega) = \det [\msys (\omega)],
\end{equation}
and can be determined using a suitable root-finding algorithm.
Setting $\omega$ equal to one specific eigenfrequency, the
corresponding eigenfunctions are first constructed on the shooting grid
$\{\xk\}$ by finding the non-trivial vector \vunk\ satisfying
eqn.~(\ref{eqn:alg-eqns}). Then, the eigenfunctions at any point in
any subinterval follow from applying eqn.~(\ref{eqn:ivp}).

Inspecting the form of \msys\ suggests that the system of equations
can be greatly simplified to
\begin{equation}
\msyscompact \, \vunkab = \vnull,
\end{equation}
where
\begin{equation} \label{eqn:unknowns-single}
\vunkab =
\begin{pmatrix}
\vy^{1} \\
\vy^{N}
\end{pmatrix}
\end{equation}
and
\begin{equation}
\msyscompact = 
\begin{pmatrix}
\mbounda     & \mnull \\
-\mfund^{N;N-1} \mfund^{N-1;N-2} \ldots \mfund^{3;2} \mfund^{2;1} & \mident \\
\mnull       & \mboundb
\end{pmatrix}.
\end{equation}
Unfortunately this approach, known as compactification, suffers from a
similar issue to single/double shooting: when evaluating the product
of fundamental solution matrices in the above expression, the columns
become numerically linearly dependent \citep[see][for a more-detailed
  discussion]{Asc1995}.

\subsection{Magnus Integrators} \label{ssec:scheme-magnus}

To evaluate the fundamental solution matrices $\mfund^{k+1;k}$ in each
of the $N-1$ subintervals, the MMS scheme builds on an approach proposed
by \citet{GabNoe1976}. These authors approximated the Jacobian matrix
of the adiabatic pulsation equations as piecewise-constant in each
subinterval (`shell' in their terminology). In the present context
this leads to a fundamental solution
\begin{equation} \label{eqn:jac-mfund}
\mfund^{k+1;k} = \exp ( \mjac \Dxk )
\end{equation}
where $\Dxk \equiv \xkp - \xk$ (a derivation of this result appears
below). This expression involves matrix exponentiation --- a topic
discussed at length by \citet{MolVanL2003}, who survey the strengths
and weaknesses of nineteen different methods. Here we focus on
eigendecomposition (their method 14), both for pedagogic purposes and
because it is adopted in the \gyre\ code
(Sec.~\ref{ssec:implement-fund}). The Jacobian matrix is written as
\begin{equation} \label{eqn:jac-eigen}
\mjac = \meigvecA \meigvalA \meigvecA^{-1},
\end{equation}
where $\meigvalA \in \field{C}^{n \times n}$ is a diagonal matrix
whose entries are the eigenvalues $\{\lambdaAi\}$ ($i=1,\ldots,n$) of
\mjac, and the columns of the matrix $\meigvecA \in \field{C}^{n
  \times n}$ comprise the corresponding eigenvectors. With this
decomposition, the fundamental solution~(\ref{eqn:jac-mfund}) becomes
\begin{equation} \label{eqn:jac-mfund-eig}
\mfund^{k+1;k} = \meigvecA \exp (\meigvalA \Dxk) \meigvecA^{-1} 
\end{equation}
where the non-zero elements of the diagonal matrix $\exp(\meigvalA
\Dxk)$ are
\begin{equation}
[\exp(\meigvalA \Dxk)]_{ii} = \exp(\lambdaAi \Dxk).
\end{equation}

An instructive physical narrative for these equations can be obtained
by substituting eqn.~(\ref{eqn:jac-mfund-eig}) into~(\ref{eqn:match}),
to yield
\begin{equation} \label{eqn:advance-y}
\vy^{k+1} = \meigvecA \exp (\meigvalA \Dxk) \meigvecA^{-1} \vyk.
\end{equation}
The matrices on the right-hand side of eqn.~(\ref{eqn:advance-y})
correspond to a sequence of operations which advance \vy\ from the
\kth\ grid point to the $k+1$ point. First, \vyk\ is projected onto a
set of basis vectors given by the rows of $\meigvecA^{-1}$. This
projection amounts to decomposing \vyk\ into contributions from $n$
independent waves. Then, the amplitudes and phases of the waves are
evolved across the subinterval by applying the diagonal matrix $\exp
(\meigvalA \Dxk)$. In evanescent zones all eigenvalues are real and
only the wave amplitudes change, whereas in propagation zones one or
more eigenvalues are complex and the wave phases also change. Finally,
the waves are projected back into physical space by the matrix
\meigvecA.

The \citet{GabNoe1976} approach can be generalized by recognizing it
as an application of a simple yet powerful theorem proposed by
\citet{Mag1954}. Subject to certain convergence criteria, the solution
to the IVP~(\ref{eqn:fund-sol}) can be written as the matrix
exponential
\begin{equation} \label{eqn:magnus-exp}
\mfund (x;x') = \exp [ \mmagnus(x;x') ],
\end{equation}
where the Magnus matrix $\mmagnus \in \field{C}^{n\times n}$ has a
series expansion whose leading terms are
\begin{multline} \label{eqn:magnus-expand}
\mmagnus(x;x') = \int_{x'}^{x} \mjac(x_{1})\,\diff x_{1} - \mbox{} \\
\frac{1}{2} \int_{x'}^{x} \left[ \int_{x'}^{x_{1}} \mjac(x_{2})\, \diff x_{2}, \mjac(x_{1}) \right] \, \diff x_{1} + \cdots
\end{multline}
(here, $[\ldots\,,\,\ldots]$ denotes the matrix
commutator). \citet{Bla2009} present a detailed review of Magnus's
theorem, covering both its mathematical underpinnings and its
practical application to solving systems of differential equations.

In the context of the MMS scheme, Magnus's theorem gives the
fundamental solution matrix within each subinterval as
\begin{equation} \label{eqn:magnus-mfund}
\mfund^{k+1;k} = \exp [ \mmagnus(\xkp;\xk) ].
\end{equation}
If the Jacobian matrix \mjac\ is independent of $x$, then all terms
but the first in the expansion~(\ref{eqn:magnus-expand}) vanish and
the Magnus matrix is simply
\begin{equation} \label{eqn:magnus-const}
\mmagnus(\xkp;\xk) = \mjac\, \Dxk.
\end{equation}
By combining eqns.~(\ref{eqn:magnus-mfund})
and~(\ref{eqn:magnus-const}) we recover the fundamental
solution~(\ref{eqn:jac-mfund}) obtained using the \citet{GabNoe1976}
approach. This constant-Jacobian case is the only one having a
closed-form expression for the Magnus matrix; however, as discussed by
\citet{Bla2009} it is relatively straightforward to construct
approximations to eqn.~(\ref{eqn:magnus-expand}) which are correct to
some specified order in the subinterval width \Dxk. Specifically, if
second-order Gauss-Legendre quadrature is used to evaluate the
integrals in the expansion~(\ref{eqn:magnus-expand}), then the Magnus
matrix becomes
\begin{equation} \label{eqn:magnus-gl2}
\mmagnus(\xkp;\xk) \approx \mjac\left(\xk + \frac{\Dxk}{2} \right) \,\Dxk + \mathcal{O}[(\Dxk)^{3}].
\end{equation}
Higher-order quadratures lead to correspondingly higher-order
approximations; in Appendix~(\ref{app:magnus}) we quote fourth- and
sixth-order accurate expressions for the Magnus matrix, also based on
Gauss-Legendre quadrature. We refer to the fundamental
solutions~(\ref{eqn:magnus-mfund}) using these expressions as the
Magnus GL\order\ integrators, where GL for stands for Gauss-Legendre
and $\order=2,4,6$ indicates the order of accuracy. These integrators
all share the useful property of not requiring Jacobian evaluations at
the subinterval endpoints, which allows them to gracefully handle
singularities at the interval boundaries.

\subsection{Commentary} \label{ssec:scheme-comment}

Compared to single/double shooting schemes, multiple shooting has the
advantage that the subinterval width \Dxk\ can always be chosen
sufficiently small that the columns of the associated fundamental
solution matrix $\mfund^{k+1,k}$ remain linearly independent, even in
the presence of an exponential dichotomy. This is because the matrix
approaches the identity matrix in the limit $\Dxk \rightarrow 0$. In
practice, this choice rarely needs to be made explicitly; small
\Dxk\ is already desirable in the interests of accuracy.

Despite being based on shooting, certain parts of the MMS scheme bear
a strong resemblance to relaxation. In particular, the staircase
structure of the system matrix~(\ref{eqn:smatrix}) also arises in
finite-difference approximations to the BVP differential equations
(see Section~\ref{ssec:background-relax}). This is no coincidence: it
is straightforward to demonstrate that any relaxation scheme can be
built from a multiple shooting scheme (and vice versa) by choosing a
suitable numerical method for the IVP~(\ref{eqn:ivp}).

In principle, eqn.~(\ref{eqn:alg-eqns}) could be solved using Castor's
method: by replacing one of the boundary conditions in the first or
last block rows with a normalization condition, the system of equations becomes
inhomogeneous and can be solved for any $\omega$. The replaced
boundary condition then serves as the discriminant function. However,
this would be an obtuse way to solve a homogeneous linear
problem, and it is little wonder that problems arise (e.g., the
singularities discussed in Section~\ref{ssec:background-relax}). The
determinant-based discriminant we propose in eqn.~(\ref{eqn:discrim})
is the natural approach, and given that the elements of \msys\ are
finite it is guaranteed to be well-behaved. \citet{Dup2001} and
\citet{Scu2008} successfully use a similar method to find trial
solutions for their inverse iteration schemes.

The MMS scheme is the first explicit application of Magnus's theorem to
stellar oscillations. However, as we demonstrate in the preceding
section the \citet{GabNoe1976} method is equivalent to shooting using
a Magnus integrator. More recently, \citet{Chr2008b} mentions that the
\adipls\ code can optionally use a similar approach (second-order
Magnus based on Newton-Cotes quadrature) to integrate the adiabatic
pulsation equations within the \citet{Cow1941} approximation. In both
papers, the authors recognize the schemes' key strength that they can
resolve solutions which vary on arbitrarily small spatial scales ---
something that fixed-stepsize IVP solvers cannot do.

%% Implementation

\section{The GYRE Code} \label{sec:implement}

\gyre\ is a new oscillation code which uses the Magnus Multiple
Shooting scheme described above to calculate the eigenfrequencies and
eigenfunctions of an input stellar model. Although \gyre\ can address
both adiabatic and non-adiabatic pulsation problems, in this paper we
focus on the adiabatic case (documented in Appendix~\ref{app:pulseqs})
because our primary goal is to introduce the MMS scheme and the code.

\gyre\ is written in Fortran 2008 with a modular architecture that
allows straightforward extension to handle more-complicated
problems. To take advantage of multiple processor cores and/or cluster
nodes it is parallelized using a combination of OpenMP
\citep{DagMen1998} and MPI \citep{Don1995}. In brief, a typical
\gyre\ run involves the following steps: first, a stellar model is
either read from file or built analytically
(Section~\ref{ssec:implement-model}), and the calculation grids are
constructed (Section~\ref{ssec:implement-grid}). A scan through
frequency space then searches for sign changes in the discriminant
$\discrim(\omega)$, which are used as initial guesses for the
discriminant roots (Section~\ref{ssec:implement-eigfreq}). After these
roots are found, the corresponding eigenfunctions are reconstructed
(Section~\ref{ssec:implement-eigfunc}). The following sections further
discuss these steps, and provide other salient implementation details.

\subsection{Stellar Model} \label{ssec:implement-model}

\gyre\ supports three classes of stellar model, each providing the
dimensionless structure coefficients $V$, $A^{*}$, $U$, $c_{1}$ and
$\Gamma_{1}$ appearing in the pulsation equations (see
Appendix~\ref{app:pulseqs}). Evolutionary models are generated by a
stellar evolution code, polytropic models are based on solutions to
the Lane-Emden equation, and analytic models rely on explicit
expressions for the structure coefficients. Both evolutionary and
polytropic models are specified on a discrete radial grid, with cubic
spline interpolation used to evaluate the structure coefficients
between grid points. Different options exist for constructing the
splines, with the monotonicity-preserving algorithm by \citet{Ste1990}
being the default.

\subsection{Grid Construction} \label{ssec:implement-grid}
 
\gyre\ offers a number of strategies for establishing the grids used
for multiple shooting (see eqn.~\ref{eqn:x-grid}) and eigenfunction
reconstruction (discussed below in
Section~\ref{ssec:implement-eigfunc}). For evolutionary and polytropic
models the grid can be cloned from the corresponding model grid, with
the option of oversampling certain subintervals. \gyre\ can also create
an ab initio grid following a variety of recipes. The simplest of
these is the `double geometric' grid with subinterval widths given by
\begin{align}
\begin{split}
\Dxk = (1 + g) \Delta \xkm \qquad k \leq M, \\
\Dxk = (1 + g) \Delta \xkp \qquad k > M.
\end{split}
\end{align}
Here, $M = N/2$ for even $N$ and $M = (N-1)/2$ for odd $N$, and the
growth factor $g$ is determined from the requirement that
\begin{equation}
\sum_{k=1}^{N-1} \Dxk = 1.
\end{equation}
The subinterval sizes at the boundaries are fixed by a user-specified
stretching parameter $s$ representing the ratio between the average
subinterval size and the boundary size; thus,
\begin{equation}
\Delta x_{1} = \Delta x_{N-1} = \frac{1}{s(N-1)}.
\end{equation}
The double geometric grid has greatest resolution near the inner and
outer boundaries --- a useful property because the components of the
Jacobian matrix typically vary fastest near these boundaries.

\subsection{Fundamental Solution Calculation} \label{ssec:implement-fund}

\gyre\ calculates the fundamental solutions with one of the Magnus
GL\order\ integrators (the order \order\ is configurable at run
time). The matrix exponential in eqn.~(\ref{eqn:magnus-mfund}) is
evaluated using eigendecomposition, and so the fundamental solution in
the \kth\ subinterval is obtained as
\begin{equation}
\mfund^{k+1;k} = \meigvecO \exp (\meigvalO) \meigvecO^{-1}.
\end{equation}
Here, \meigvalO\ and \meigvecO\ are the eigenvalue and eigenvector
matrices of the Magnus matrix $\mmagnus(\xkp;\xk)$, itself taken from
one of eqns.~(\ref{eqn:magnus-gl2}), (\ref{eqn:magnus-gl4})
or~(\ref{eqn:magnus-gl6}) for the GL2, GL4 or GL6 integrators,
respectively. The eigendecomposition is implemented with calls to the
\lapack\ linear algebra library \citep{And1999}; OpenMP directives are
used to distribute the work for the $N-1$ subintervals across multiple
processor cores.

The eigendecomposition can fail if $\mmagnus(\xkp;\xk)$ lacks a
complete set of linearly independent eigenvectors \citep[i.e., the
  matrix is defective; see][]{GolVanL1996}. In such cases one of the
alternative matrix exponentiation methods discussed by
\citet{MolVanL2003} must be used. In practice we have never
encountered this situation; nevertheless, as a precaution \gyre\ is
configured to abort with an error when it detects a defective or
near-defective Magnus matrix.

\subsection{Determinant Evaluation} \label{ssec:implement-det}

To evaluate the determinant of the system matrix \msys, \gyre\ first
constructs the LU factorization
\begin{equation} \label{eqn:lu-factor}
\msys = \mL \mU,
\end{equation}
where \mL\ is a lower-triangular matrix with unit diagonal elements
and \mU\ is an upper triangular matrix. The determinant of a
triangular matrix is simply the product of its diagonal elements, and
it therefore follows that
\begin{equation} \label{eqn:det-eval}
\det(\msys) = \prod_{j=1}^{N n} \mU_{jj}.
\end{equation}

\gyre\ undertakes the factorization~(\ref{eqn:lu-factor}) using the
structured algorithm described by Wright
\citeyearpar{Wri1992,Wri1994}, which is specifically targeted at
matrices arising in multiple shooting schemes. The cyclic reduction
version of the algorithm is implemented because it produces the same
results whether run serially or in parallel. Block-row pairs are
distributed across multiple processor cores using OpenMP directives,
and then factorized with calls to the \lapack\ and
\blas\ \citep{Bla2002} libraries. We explored an MPI version of the
algorithm for use on clusters, but found that inter-node communication
overhead produces poor performance on systems larger than a few tens
of nodes. As the dimension of \msys\ becomes large the determinant
risks overflowing the computer floating-point range; therefore,
\gyre\ evaluates eqn.~(\ref{eqn:det-eval}) with extended-range
floating point arithmetic, build on the object-oriented capabilities
of recent Fortran dialects.

\subsection{Eigenfrequency Searching} \label{ssec:implement-eigfreq}

\gyre\ searches for eigenfrequencies within a user-specified frequency
interval $\omegaa \le \omega \le \omegab$ by first evaluating the
discriminant $\discrim(\omega)$ at $n_{\omega}$ points distributed
within this interval. A change in the sign of $\discrim(\omega)$
between an adjacent pair of points signals that a root is bracketed,
and the pair is passed as starting guesses to a root-finding routine
based on the algorithm described by \citet{Bre1973}.

To leverage multi-node clusters \gyre\ parallelizes the initial
discriminant evaluations and subsequent root searches with calls to
the MPI library. This is in addition to the OpenMP parallelization
described above for the eigendecompositions and LU factorization.

\subsection{Eigenfunction Reconstruction} \label{ssec:implement-eigfunc}

For each eigenfrequency found as a root of $\discrim(\omega)$,
\gyre\ reconstructs the corresponding eigenfunctions on the shooting
grid $\{\xk\}$ by solving the algebraic
system~(\ref{eqn:alg-eqns}). The LU
factorization~(\ref{eqn:lu-factor}) reduces this system to
\begin{equation} \label{eqn:lu-sol}
\mU \, \vunk = \vnull.
\end{equation}
Because the upper triangular matrix \mU\ is singular when $\omega$ is
an eigenfrequency, one of its diagonal elements --- say, $\mU_{j'j'}$
--- must be zero to within numerical uncertainties. The elements of
the solution vector can then be written as
\begin{equation} \label{eqn:vunk-sol}
\vunk_{j} = 
\begin{cases}
- [ \mUred^{-1} \vunkred ]_{j} & j < j', \\
  1                           & j = j', \\
  0                           & j > j',
\end{cases}
\end{equation}
where the square matrix \mUred\ is formed from the first $j'-1$ rows
and columns of \mU, and the vector \vunkred\ is formed from the first
$j'-1$ elements of the $j'$ column of \mU. Because \mUred\ is upper
triangular, the product $\mUred^{-1} \vunkred$ is evaluated trivially
by back-substitution \citep[e.g.,][]{GolVanL1996}. With
\vunk\ determined in this way, $\{\vyk\}$ can be unpacked using
eqn.~(\ref{eqn:vunk}).

Non-trivial solutions to the pulsation equations cannot completely
vanish at any point (otherwise, they would vanish everywhere). This
means that the zero element $\mU_{j'j'}$ should be located in the
bottom-right corner of the matrix, such that $j' > (N-1)
n$. \gyre\ explicitly tests whether this condition is met, and flags
violations to indicate a problem with the solution. Our experience has
been that these violations arise when the grid spacing is too large in
one or more subintervals, preventing the Magnus
expansion~(\ref{eqn:magnus-expand}) from converging. The fix is
invariably to reduce \Dxk\ by increasing $N$.

Once the eigenfunctions are obtained on the shooting grid,
\gyre\ evaluates them on the separate reconstruction grid using a
secant-line approximation to the Magnus matrix within each
subinterval. In the \kth\ subinterval this is
\begin{equation}
\mmagnus(x;\xk) \approx \wk(x) \mmagnus(\xkp;\xk),
\end{equation}
where the weight function
\begin{equation}
\wk(x) \equiv \frac{x - \xk}{\Dxk} 
\end{equation}
varies between 0 ($x=\xk$) and 1 ($x=\xkp$). The solution at any point
in the subinterval is then efficiently calculated as
\begin{equation}
\vy(x) = \meigvecO \exp [ \meigvalO \, \wk(x) ] \meigvecO^{-1} \vyk,
\end{equation}
where \meigvalO\ and \meigvecO\ are the same eigenvalue and
eigenvector matrices obtained during the fundamental solution
construction (cf. Sec.~\ref{ssec:implement-fund}); no further
eigendecomposition is required.

Eigenfunctions resulting from this procedure are $C^{\infty}$
continuous within subintervals and are $C^{0}$ continuous at the
edges. As final steps \gyre\ normalizes the eigenfunctions to have a
mode inertia $\inertia = M R^{2}$ \citep[see][their
  eqn.~3.139]{Aer2010}, and then classifies them in the standard
\citet{Eck1960}--\citet{Scu1974}--\citet{Osa1975} (ESO) scheme by
enumerating the acoustic- and gravity-wave winding numbers \npr\ and
\ngr. For dipole modes the ESO scheme can fail in certain
circumstances \citep[see, e.g.,][]{ChrMul1994}, and so \gyre\ instead
uses the modified scheme developed by \citet{Tak2006}.

%% Calculations

\section{Calculations} \label{sec:calc}

\subsection{Eigenfrequencies of the $n=0$ Polytrope} \label{ssec:calc-poly}

\begin{figure}
\includegraphics{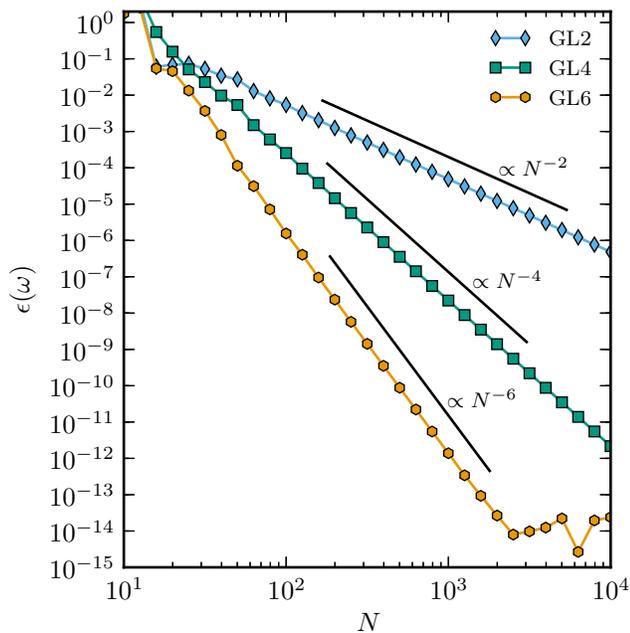}
\caption{The absolute error in the dimensionless eigenfrequency of the
  dipole \pone\ mode of the $n=0$ polytrope, plotted as a function of
  the number of grid points. The three curves correspond to \gyre's
  GL2, GL4 and GL6 Magnus integrators, while the thick lines show the
  corresponding asymptotic scalings $\error(\omega) \propto N^{-2}$,
  $N^{-4}$ and $N^{-6}$, respectively.} \label{fig:poly}
\end{figure}

As an initial test of \gyre\ and the underlying MMS scheme, we
calculate radial and non-radial eigenfrequencies of the $n=0$
polytrope model with $\Gamma_{1} = 5/3$ (the so-called homogeneous
compressible sphere). Because exact expressions exist for these
frequencies \citep{Pek1938} this exercise allows an assessment of how
the code and the scheme perform as the resolution of the shooting grid
is varied.

Figure~\ref{fig:poly} illustrates typical results, plotting the
absolute error $\error(\omega) \equiv |\omega - \omegaex|$ in the
eigenfrequency (with \omegaex\ being the exact value) as a function of
the number of grid points $N$, for the $\ell=1$ mode with
$(\npr,\ngr)=(1,0)$ (traditionally labeled the dipole
\pone\ mode). The three curves show data from runs using the GL2, GL4
and GL6 Magnus integrators; in all cases the shooting grid is the
double geometric grid described in Section~\ref{ssec:implement-grid}
with stretching parameter $s = 10^{3}$.

The figure clearly reveals that the eigenfrequency error follows an
asymptotic scaling $\propto N^{-2}$, $N^{-4}$ and $N^{-6}$ for the
GL2, GL4 and GL6 integrators, respectively. This is the expected
behavior: the GL\order\ integrator is \order'th order accurate,
leading to fundamental solutions (cf. eqn.~\ref{eqn:magnus-mfund})
with an error scaling as $\Dx^{\order+1}$. Accumulated over the $N-1$
subintervals the global error of the shooting scheme is then $\error
\sim (N-1) \Dx^{\order+1} \sim N^{-\order}$ (where we have used $\Dx
\sim N^{-1}$), which is the scaling seen in the figure. (For $N
\gtrsim 2 \times 10^{3}$ the GL6 integrator departs from the
asymptotic behavior described, because numerical rounding becomes the
dominant contributor toward the error).

Results for other radial and non-radial modes are comparable to those
shown in the figure for the dipole \pone\ mode. This confirms that the
MMS scheme with the Magnus GL2, GL4 and GL6 integrators yields
eigenfrequencies whose departures from exact values scale as the
inverse second, fourth and sixth power of the grid size.

\subsection{Inter-Code Comparison with ESTA Model M4k} \label{ssec:calc-esta}

\begin{figure*}
\includegraphics{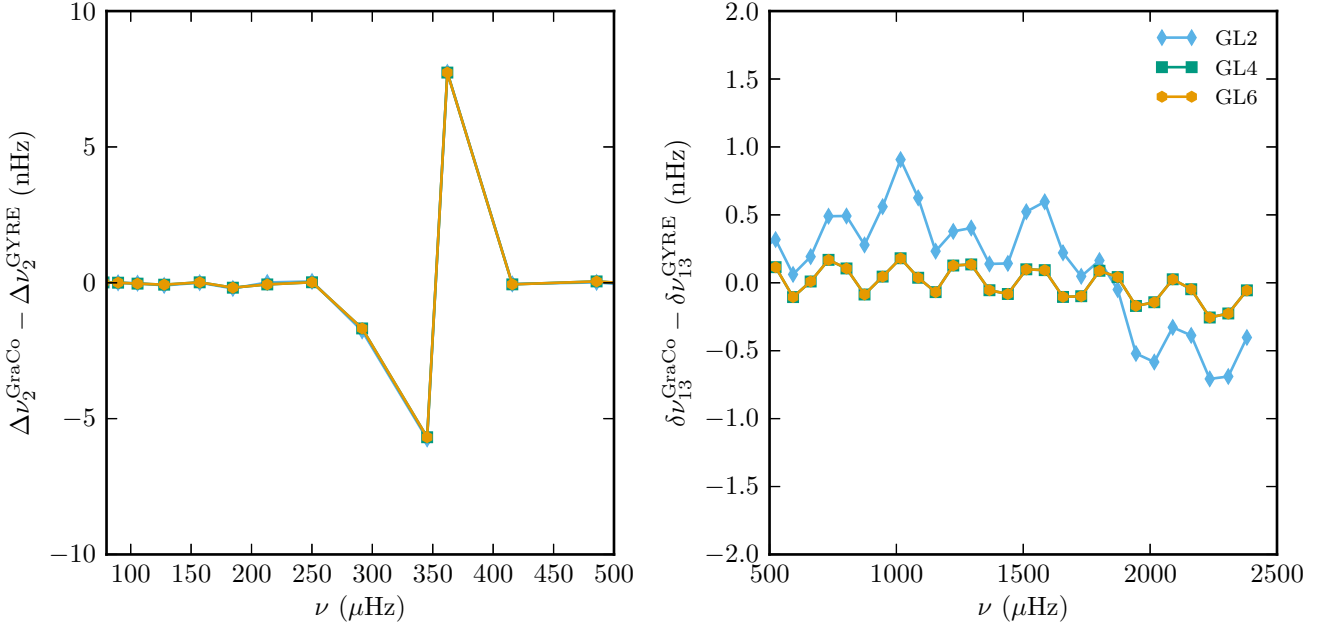}
\caption{Differences between the \graco\ and \gyre\ $\ell=2$ large
  frequency separations (left) and $\ell=1,3$ small frequency
  separations (right) of the M4k model, plotted as a function of
  linear frequency. The three curves correspond to \gyre's GL2, GL4
  and GL6 Magnus integrators; in the left-hand panel, all three curves
  sit atop each other, while in the right-hand panel the GL4 and GL6
  curves overlap. The panels should be compared against Figs.~12
  and~15, respectively, of \citet{Moy2008}} \label{fig:esta}
\end{figure*}

\begin{table}
\begin{tabular}{cccccccccc}
        & \multicolumn{3}{c}{GL2}  & \multicolumn{3}{c}{GL4}  & \multicolumn{3}{c}{GL6} \\
 $\ell$ & Low & Med. & High & Low & Med. & High & Low & Med. & High \\ \hline
0 & --- & 0.44 & 3.39 & --- & 0.39 & 0.49 & --- & 0.39 & 0.48 \\
1 & 0.67 & 3.61 & 4.39 & 0.64 & 3.39 & 0.64 & 0.64 & 3.39 & 0.64 \\
2 & 1.15 & 7.83 & 3.90 & 1.10 & 7.72 & 0.38 & 1.10 & 7.72 & 0.38 \\
3 & 1.63 & 13.11 & 3.55 & 1.57 & 13.12 & 0.34 & 1.57 & 13.12 & 0.34 \\
\end{tabular}
\caption{Maximum absolute differences, in \nHz, between the
  \graco\ and \gyre\ linear frequencies for $l=0,\ldots,3$ modes of
  the M4k model. Values are tabulated for each of \gyre's integrators
  and for the same low ($20\,\muHz \leq \nu \leq 80\,\muHz$), medium
  ($80\,\muHz \leq \nu \leq 500\,\muHz$) and high ($500\,\muHz \leq
  \nu \leq 2\,500\,\muHz$) frequency regions adopted for discussion
  purposes by \citet{Moy2008}.} \label{tab:esta}
\end{table}

As a second verification exercise, we use \gyre\ to calculate
eigenfrequencies of the `M4k' model described by \citet{Moy2008}. This
model was produced with the \astec\ stellar evolution code
\citep{Chr2008a} and represents a 1.5\,\Msun\ star at an age
$1.35\,\Gyr$, about half-way through its main sequence evolution; it
was adopted by \citet{Moy2008} as the basis for their comprehensive
comparison of oscillation codes\footnote{All mentioned in
 Section~\ref{sec:background}.} from nine different research groups
participating in the \corot\ \textit{Evolution and Seismic Tools}
activity \citep[ESTA; see][]{Leb2008}.  \gyre's shooting grid is
cloned from the model grid without any oversampling (see
Section~\ref{ssec:implement-grid}), and the gravitational constant is
set to the same value $G=6.6716823 \times 10^{-8}
\cm^{3}\,\gram^{-1}\,\second^{-2}$ adopted by \citet{Moy2008}. As in
the preceding section, we perform separate runs using the GL2, GL4 and
GL6 Magnus integrators.

Table~\ref{tab:esta} compares the \gyre\ linear eigenfrequencies
against those obtained with \graco, which was used as the reference
code in the \citet{Moy2008} study. Across the range $20\,\muHz \leq
\nu \leq 2\,500\,\muHz$ considered by these authors the absolute error
between the \gyre\ and \graco\ frequencies is $\la 4\,\nHz$ for radial
modes, rising to $\la 14\,\nHz$ in the $\ell=3$ case. These values are
comparable to the frequency differences found by \citet{Moy2008}
between \graco\ and the other oscillation codes.  The GL4 and GL6
integrators produce almost identical results, indicating that the
frequency differences between them are much smaller than their
differences with \graco.

To further illustrate the comparison between \gyre\ and \graco,
Fig.~\ref{fig:esta} plots the differences between the $\ell=2$ large
separations \Dnu\ and the $\ell=1,3$ small separations \dnu, both as a
function of frequency \citep[see, e.g.,][for a definition and
  discussion of these asteroseismic parameters]{Aer2010}. These two
plots are intended for direct comparison against the middle panels of
Figures~12 and 15, respectively, of \citet{Moy2008}. They confirm that
\gyre\ is in good agreement with the other oscillation codes. (The
zig-zag feature at $\nu \approx 350\,\muHz$ in the left panel of
Fig.~\ref{fig:esta} is also seen when comparing \graco\ against other
codes, and therefore is not due to \gyre).

\begin{figure}
\includegraphics{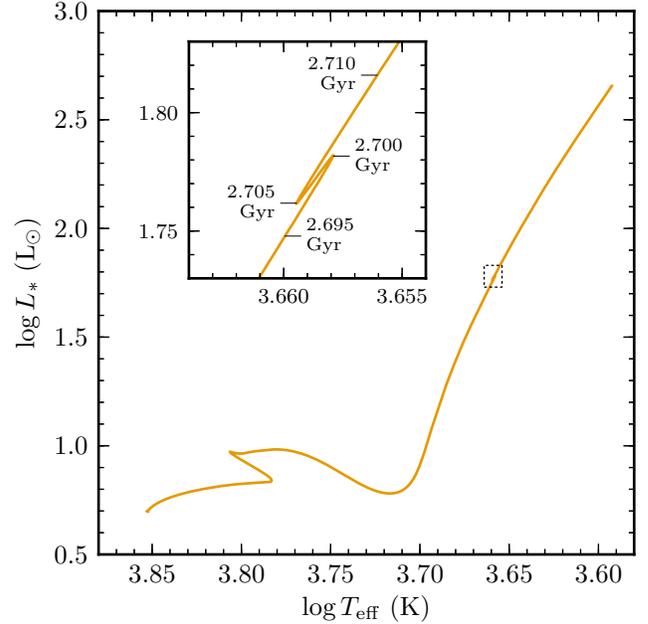}
\caption{The evolutionary track of the 1.5\,\Msun\ star plotted in the
  HRD. The inset magnifies the RGB bump phase (shown in the
  main diagram by the dotted rectangle), where the star's luminosity
  growth undergoes a temporary reversal; tick marks indicate the
  stellar age.} \label{fig:rgb-hrd}
\end{figure}

\subsection{Asteroseismology through the RGB Bump} \label{ssec:calc-rgb}

\begin{figure*}
\includegraphics{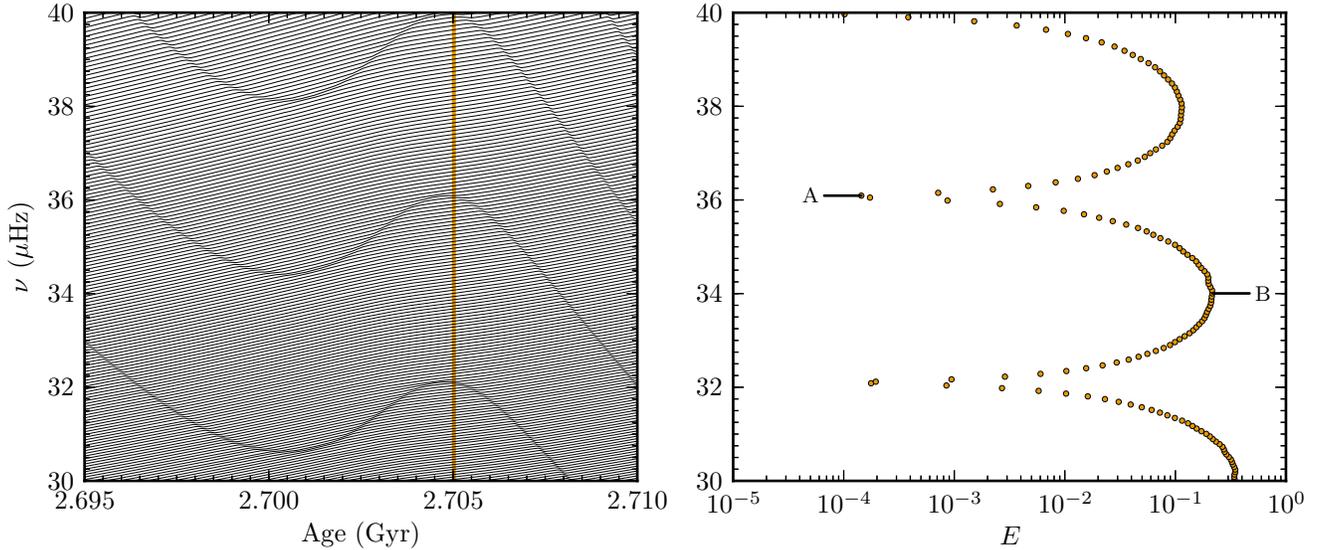}
\caption{Linear eigenfrequencies of dipole modes of the
  1.5\,\Msun\ model plotted as a function of stellar age (left),
  together with the normalized mode inertia of the 2.705\,\Gyr\ model
  (right; also highlighted in the left panel by the vertical
  line). The differential inertia of the modes labeled A and B is
  shown in
  Fig.~\ref{fig:rgb-eigfuncs}.} \label{fig:rgb-freq-inertia}
\end{figure*}

As a `first science' experiment using \gyre\ we now explore how the
seismic properties of a 1.5\,\Msun\ star change as it passes through
the so-called red giant branch (RGB) bump. During this evolutionary
phase the hydrogen burning shell encounters the abundance
discontinuity left by the convective envelope during first dredge-up,
causing a temporary reversal in the star's luminosity growth as it
ascends the RGB. This reversal shows up in the Hertzsprung-Russell
diagram (HRD) as a narrow zig-zag in a single star's evolutionary
track, and it causes a bump in the luminosity function of cluster
members on the RGB, hence the name \citep[see][for a more detailed
  discussion]{Sal2002}. Our decision to focus on the red bump has two
motivations: on the one hand RGB stars are an area of especial recent
interest (see Section~\ref{sec:intro}), and on the other they are a
challenge for any oscillation code to model due to the extremely short
spatial scale of eigenfunctions in their radiative cores.

Models for the 1.5\,\Msun\ star spanning the bump phase are obtained
by running the \texttt{1.5M\_with\_diffusion} test-suite calculation
of the \mesastar\ stellar evolution code, revision 4930
\citep[see][]{Pax2011,Pax2013}. The resulting track in the HRD is
plotted in Fig.~\ref{fig:rgb-hrd}, with the inset magnifying the RGB
bump phase. For each of the $\sim 300$ models spanning this phase we
use \gyre\ to find $\ell=1$ modes in the frequency range $30\,\muHz
\leq \nu \leq 40\,\muHz$, chosen to loosely correspond to the
frequency of maximum power \numax\ predicted by the standard scaling
relation for solar-like oscillations
\citep[e.g.,][]{Bro1991,KjeBed1995}. Fig.~\ref{fig:rgb-freq-inertia}
illustrates results from this exercise, plotting linear frequencies as
a function of model age. The figure also shows the frequency
dependence of the normalized mode inertia \norminertia\ \citep[as
  defined by][their eqn.~3.140; not to be confused with the
  un-normalized inertia \inertia]{Aer2010} for a single model with an
age $\approx 2.705\,\Gyr$ which places it near the minimum luminosity
encountered during the bump phase.

\begin{figure*}
\includegraphics{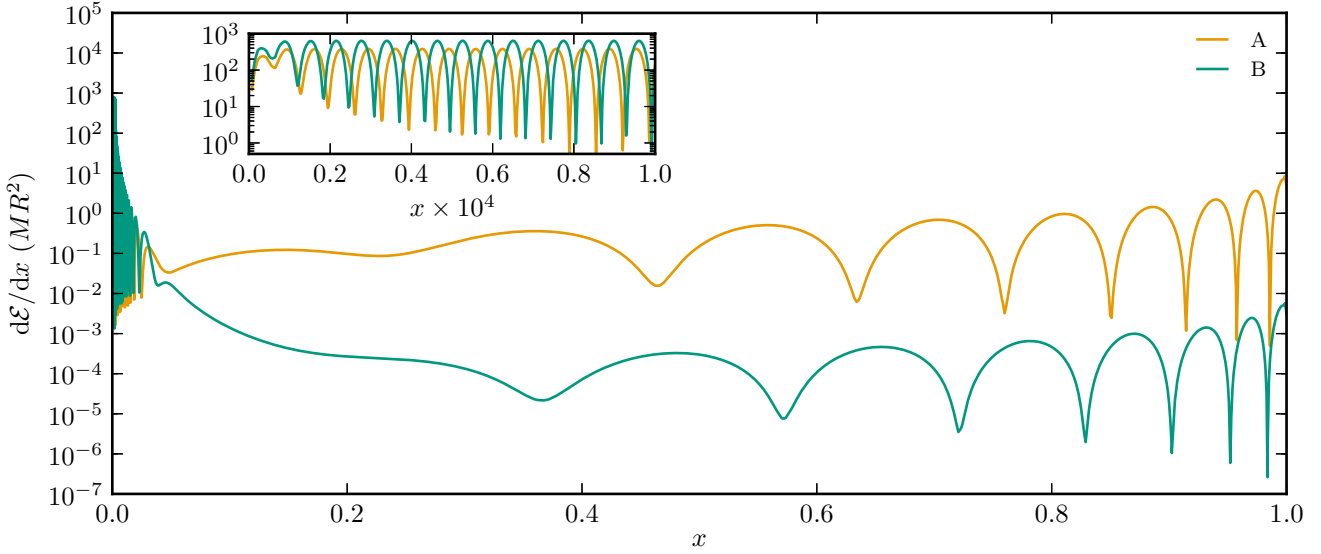}
\caption{The differential inertia of the two modes labeled
  in the right-hand panel of Fig.~\ref{fig:rgb-freq-inertia}, plotted
  as a function of fractional radius. The formal classifications are
  $(\npr,\ngr)=(8,476)$ for mode A and $(\npr,\ngr)=(7,505)$ for mode
  B. The inset magnifies the centermost region, illustrating the very
  small spatial scale of the modes there.} \label{fig:rgb-eigfuncs}
\end{figure*}

\begin{figure*}
\includegraphics{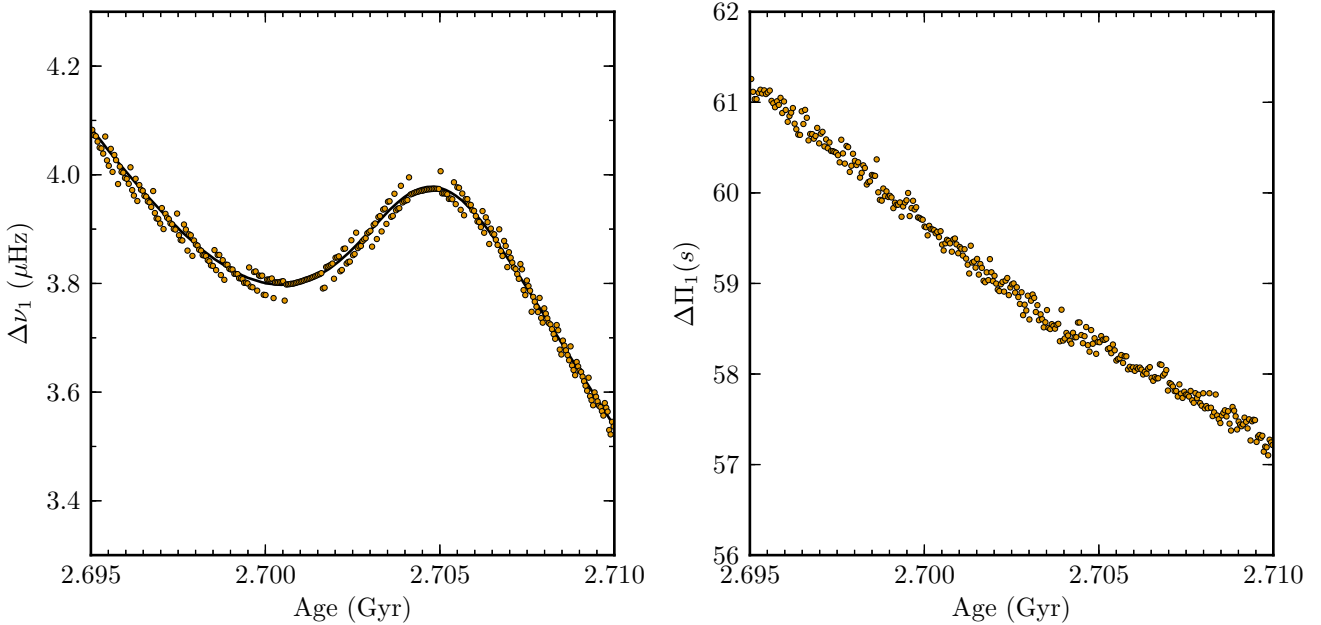}
\caption{The p-mode large frequency separation (left) and g-mode
  period separation (right) of dipole modes of the 1.5\,\Msun\ model,
  plotted as a function of stellar age. The large separation is
  measured between the minima of the normalized inertia arising from
  avoided crossings with the $\npr=7$ and $\npr=8$ envelope p-modes,
  while the period separation is taken as the maximum period spacing
  of modes situated between these two minima. The solid curve in the
  left panel follows the scaling $\Dfreq \propto \tdyn^{-1}$, and is
  normalized to pass through the initial data
  point.} \label{fig:rgb-seps}
\end{figure*}

The figure depicts many of the features characteristic to red-giant
oscillations. The frequency spectrum is dominated by a dense forest of
g-modes with radial orders $n \equiv \npr-\ngr$ in the range $-577
\leq n \leq -401$. These modes are trapped in the radiative interior
of the star where the \BV\ frequency is large. A small subset of the
modes have frequencies close to those of envelope p-modes, and
coupling between the two leads to the distinctive pattern of avoided
crossings \citep{Aiz1977} displayed in the left-hand panel of the
figure. During an avoided crossing, the greatly enhanced mode
amplitude in the low-density stellar envelope leads to a much-reduced
normalized inertia, as can be seen in the right-hand panel. It is
these low-inertia modes which dominate the observed frequency spectra
of RGB stars, as they are easiest to excite to measurable amplitudes
by stochastic processes \citep[see, e.g.,][]{ChaMig2013}.

To further illustrate the change in mode properties during an avoided
crossing, Fig.~\ref{fig:rgb-eigfuncs} plots the differential inertia
$\diff \inertia/\diff x$ (which is proportional to the kinetic energy
density) as a function of fractional radius $x$ for the modes labeled
`A' and `B' in the right-hand panel of
Fig.~\ref{fig:rgb-freq-inertia}. Mode A is involved in an avoided
crossing and has an appreciable amplitude in both core and envelope.
In contrast Mode B is confined to the radiative core and has a
negligible amplitude at the surface, accounting for its much enhanced
normalized inertia compared to mode A. For both modes the radial
wavelength in the radiative interior is very short due to the large
\BV\ frequency there, leading to the highly oscillatory behavior
(well-resolved by \gyre) seen in the inset of the figure.

Returning to Fig.~\ref{fig:rgb-freq-inertia}, the evolution through
the RGB bump phase reveals itself by a temporary increase in the
otherwise-decreasing frequencies of the avoided crossings. This is a
direct consequence of contraction of the star's envelope during the
bump luminosity reversal, which shortens the sound crossing time and
therefore elevates p-mode frequencies. The frequencies of g-modes are
largely unaffected, as the radiative interior of the star changes only
slowly during bump passage.

The same general behavior can also be seen in Fig~\ref{fig:rgb-seps},
which plots asteroseismic observables --- the p-mode large frequency
separation $\Dfreq_{1}$ and the g-mode period separation $\dP_{1}$ ---
as a function of stellar age for the same $\ell=1$ modes. The
frequency separations closely follow the scaling $\Dfreq \propto
\tdyn^{-1}$ predicted by asymptotic relations \citep[e.g.,][]{Aer2010},
where $\tdyn \equiv \sqrt{R^{3}/G M}$ is the star's dynamical
timescale. The gradual decrease in the period separation arises due to
the growing mass and shrinking radius of the degenerate helium core,
which together raise the gravitational acceleration and hence
\BV\ frequency there.

Motivated by this analysis, we can speculate whether the RGB bump
manifests itself in asteroseismic observables. In their presentation
of initial \kepler\ observations, \citet{Kal2010} discuss a distinct
subpopulation of RGB stars which they identify as bump stars. However,
it might be argued that these stars (seen, e.g., as the `B' feature in
their Fig. 9b) are simply an extension of the core-helium-burning red
clump stars to lower effective temperatures. To explore this issue
further, it is necessary first to disentangle the RGB and the red
clump. As demonstrated by \citet{Bed2011} this separation can be
achieved on the basis of measured period separations, which are now
becoming available for large numbers of stars \citep{Ste2013}.

\subsection{Parallel Scaling} \label{ssec:calc-parallel}

\begin{figure*}
\includegraphics{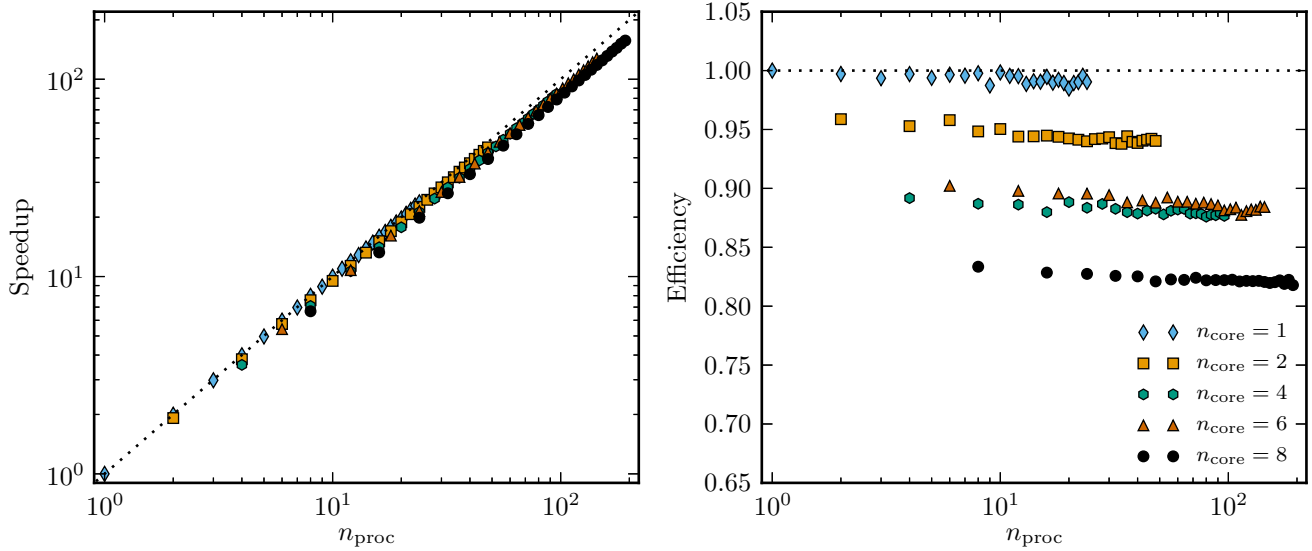}
\caption{The speedup (left) and efficiency (right) of the root
  bracketing calculations, plotted as a function of the total number
  of processors for different combinations of cores and nodes. Each
  point is based on an average over five separate runs. The dotted
  lines show the ideal linear speedup and unity efficiency
  cases.} \label{fig:timing}
\end{figure*}

To explore how the performance of \gyre\ scales on parallel
architectures, we measure the execution time $T$ of the initial root
bracketing calculations (see Section~\ref{ssec:implement-eigfreq}) for
$\ell=1$ modes of the M4k model introduced in
Section~\ref{ssec:calc-esta}. These calculations are undertaken on a
cluster of 24 nodes, each containing two 4-core AMD Opteron
processors, networked on an Infiniband switched
fabric. Figure~\ref{fig:timing} illustrates results from calculations
using different numbers of nodes \nnode\ and cores per node \ncore,
plotting the speedup and efficiency against the total number of
processors $\nproc \equiv \nnode \cdot \ncore$. The speedup
$T(1)/T(\nproc)$ measures the overall performance of the code relative
to the single-processor case, while the efficiency $T(1)/[\nproc
  T(\nproc)]$ indicates how effectively individual processors are
utilized, again relative to the single-processor case.

As discussed in Section~\ref{sec:implement}, \gyre\ implements a
hybrid approach to parallelization: OpenMP directives allow multiple
cores to participate in the construction and subsequent LU
factorization of a single system matrix, while MPI calls allow cluster
nodes to evaluate multiple discriminants concurrently. The figure
confirms that this approach is largely successful, with the speedup
increasing monotonically with \nproc. While a modest decline in
efficiency can be seen as the number of cores per node grows, this
isn't much cause for concern when running \gyre\ on today's commonly
available multi-core architectures. That said, to take full advantage
of next-generation architectures such as Intel's many-core Xeon Phi
co-processor (used in the NSF Stampede cluster at the Texas Advanced
Computing Center) it will be necessary to further improve the code's
efficiency, presumably via increased OpenMP parallelization.

%% Discussion & summary

\section{Discussion \& Summary} \label{sec:discuss}

In the preceding sections we introduce a new Magnus Multiple Shooting
scheme for solving linear homogeneous boundary value problems
(Section~\ref{sec:scheme}), together with an oscillation code
\gyre\ that implements this scheme to calculate eigenfrequencies and
eigenfunctions of stellar models
(Section~\ref{sec:implement}). Initial test calculations indicate that
the code is accurate, robust, and makes efficient use of computational
resources (Section~\ref{sec:calc}).

\gyre\ debuts in an arena which is already well populated with
oscillation codes (cf. Section~\ref{sec:background}). However, of all
these codes only \adipls\ is freely available, and it is restricted to
adiabatic pulsation --- clearly not an optimal arrangement given the
data analysis challenges now facing the field
(Section~\ref{sec:intro}). We are therefore pleased to make the
\gyre\ source code open\footnote{See
  \url{http://www.astro.wisc.edu/~townsend/gyre/}} for use and
distribution under the GNU General Public License. Our hope is that a
community of practice \citep[e.g.,][]{Tur2013} will arise around the
code, bringing together users and developers to shape the code's
future evolution in ways that best serve the field and its
participants.

As we mention in Section~\ref{sec:implement}, although the present
paper focuses primarily on adiabatic pulsation, \gyre\ can also
address non-adiabatic problems. This capability requires a few minor
adjustments to the MMS scheme, already implemented in GYRE, which we
plan to describe in detail in a forthcoming science-oriented
paper. Looking further into the future, we intend to extend \gyre\ to
include the effects of stellar rotation --- first within the
traditional approximation \citep[e.g.,][]{Tow2005}, and then using the
spherical harmonic expansion approach pioneered by \citet{DurSku1968}
and recently adopted by various groups (e.g., \citealt{Lee2001};
\citealt{Ree2006}; \citealt{Oua2012}). The expansion approach results in
BVPs with large numbers of unknowns ($4 h$ in the adiabatic
approximation when $h$ spherical harmonics are used), and will
therefore be a particularly appropriate target for testing the
robustness and performance scalability of the MMS scheme and \gyre.

Alongside these code development activities, we plan to interface
\gyre\ with the \mesastar\ evolution code. \gyre\ can already
natively read models produced by
\mesastar\ (cf. Section~\ref{ssec:calc-rgb}); the next step is to wrap
\gyre\ in a callable interface and integrate it into \mesastar's
asteroseismic module. This will open up the possibility of
large-scale, automated adiabatic and non-adiabatic asteroseismic
analyses, in turn facilitating investigation of issues such as core
rotation in RGB stars \citep[e.g.,][]{Mos2012}, instability strips in
white dwarfs \citep[e.g.,][]{FonBra2008}, and the surprising incidence
of opacity-driven oscillations in low-metallicity environments
\citep[e.g.,][]{Sal2012}.

%% Acknowledgements

\thanks

\section*{Acknowledgments}

We acknowledge support from NSF awards AST-0908688 and AST-0904607 and
NASA award NNX12AC72G. We are very grateful to Pieter Degroote, Mike
Montgomery, Dennis Stello, Chris Cameron and Bill Paxton for their
help and support in testing the \gyre\ code; to Andy Moya for sharing
ESTA data with us; to the GFORTRAN compiler development team for their
fantastic responsiveness to support requests; and to the anonymous
referee for constructive comments. We also thank the Kavli Institute
for Theoretical Physics at the University of California-Santa Barbara
for the very stimulating \emph{Asteroseismology in the Space Age}
program which helped inspire \gyre. This research has made use of
NASA's Astrophysics Data System.

%% References

\bibliographystyle{mn2e}
\bibliography{gyre}

%% Appendices

\appendix

\section{Pulsation Equations} \label{app:pulseqs}

This appendix briefly summarizes the pulsation BVP solved by \gyre\ in
the adiabatic case. The independent variable is the fractional radius
$x = r/R$, with $r$ the radial coordinate and $R$ the stellar radius,
while the components of the dependent variable vector \vy\ are
\begin{multline}
y_{1} = \frac{\xi_{r}}{r} x^{2-\ell}, \qquad
y_{2} = \frac{1}{g r} \left( \frac{p'}{\rho} + \Phi' \right) x^{2-\ell}, \\
y_{3} = \frac{1}{gr} \Phi' x^{2-\ell}, \qquad
y_{4} = \frac{1}{g} \frac{\diff \Phi'}{\diff r} x^{2-\ell}.
\end{multline}
Here, the symbols have the same meaning as in \citet{Unn1989};
specifically, $\xi_{r}$ is the radial displacement perturbation and
$p'$ and $\Phi'$ are the Eulerian perturbations to the pressure and
gravitational potential, respectively. These definitions mirror the
dimensionless variables introduced by \citet{Dzi1971}, except that we
introduce the scaling $x^{2-\ell}$ to ensure that the variables
approach constant values at the origin $x=0$ --- desirable behavior
from a numerical perspective.

Given the definitions above, the differential equations governing
linear, adiabatic non-radial oscillations can be written in the
canonical form~(\ref{eqn:diff-eqns}) with a Jacobian matrix
\begin{equation} \label{eqn:pulseqs-jac}
\mjac = x^{-1} \mjacnorm,
\end{equation}
where
\begin{equation} \label{eqn:pulseqs-jacnorm}
\mjacnorm = 
\begin{pmatrix}
\frac{V}{\Gamma_{1}} - 1 - \ell   &\frac{\ell(\ell+1)}{c_{1} \omega^{2}} - \frac{V}{\Gamma_{1}} & \frac{V}{\Gamma_{1}}                & 0 \\
c_{1}\omega^{2}-A^{*} & A^{*} - U +3 - \ell                         & - A^{*}            & 0 \\
0                    & 0                                          & 3 -U - \ell        & 1 \\
U A^{*}              & \frac{U V}{\Gamma_{1}}                      & \ell(\ell+1)-\frac{U V}{\Gamma_{1}} & -U + 2 - \ell \\
\end{pmatrix}.
\end{equation}
The dimensionless oscillation frequency $\omega$ is related to the
linear frequency $\nu$ via
\begin{equation} \label{eqn:freqs}
\omega = 2 \pi \nu \sqrt{\frac{R^{3}}{GM}},
\end{equation}
and the other variables again have the same meaning as in
\citet{Unn1989}.

The requirement that solutions remain regular at the center leads to
the inner boundary conditions
\begin{equation} \label{eqn:bc-inner}
\mbounda = 
\begin{pmatrix}
c_{1} \omega^{2} & -\ell & 0 & 0 \\
0 & 0 & \ell & -1 \\
\end{pmatrix},
\end{equation}
evaluated at $x = \xa = 0$.
Likewise, the requirement that the Lagrangian pressure perturbation
$\delta p$ vanishes at the stellar surface, and that $\Phi'$ vanishes
at infinity, leads to the outer boundary conditions
\begin{equation} \label{eqn:bc-outer}
\mboundb =
\begin{pmatrix}
1 & -1 & 1 & 0 \\
U & 0 & \ell+1 & 1
\end{pmatrix}
\end{equation}
evaluated at $x = \xb = 1$. \gyre\ offers the option of outer boundary
conditions based on more-sophisticated treatments of the stellar
atmosphere; these include the prescriptions by \citet{Dzi1971} and
\citet{Unn1989}. However, for all calculations presented in
Sec.~\ref{sec:calc} the zero-$\delta p$ condition incorporated in
eqn.~(\ref{eqn:bc-outer}) is adopted.

\section{Magnus Matrices} \label{app:magnus}

For convenience, this section presents expressions for the Magnus
matrices \citep[taken from][]{Bla2009} which are used by \gyre's GL4
and GL6 Magnus integrators.

\subsection{GL4 Magnus Integrator}

Using a fourth-order Gauss-Legendre quadrature, the Magnus matrix in
the \kth\ subinterval is approximated as
\begin{equation} \label{eqn:magnus-gl4}
\mmagnus(\xkp;\xk) \approx \malpha_{4,1} - \frac{1}{12} [ \malpha_{4,1}, \malpha_{4,2} ] + \mathcal{O}[(\Dxk)^{5}].
\end{equation}
Here,
\begin{equation}
\malpha_{4,1} = \frac{\Dxk}{2} (\mjac_{1} + \mjac_{2}), \qquad 
\malpha_{4,2} = \Dxk \sqrt{3} (\mjac_{2} - \mjac_{1}),
\end{equation}
and $\mjac_{i} \equiv \mjac(\xk_{i})$ ($i=1,2$) are the Jacobian
matrices evaluated at the two Gauss-Legendre nodes within the
subinterval,
\begin{equation}
\xk_{1} = \xk + \left( \frac{1}{2} - \frac{\sqrt{3}}{6} \right) \Dxk, \qquad
\xk_{2} = \xk + \left( \frac{1}{2} + \frac{\sqrt{3}}{6} \right) \Dxk.
\end{equation}
Note that the above expression for $\malpha_{4,2}$ corrects an error
in eqn.~(253) of \citet{Bla2009}.

\subsection{GL6 Magnus Integrator}

Using a sixth-order Gauss-Legendre quadrature, the Magnus matrix in
the \kth\ subinterval is likewise approximated as
\begin{multline} \label{eqn:magnus-gl6}
\mmagnus(\xkp;\xk) \approx \malpha_{6,1} + \frac{1}{12} \malpha_{6,3} + \mbox{} \\ 
\frac{1}{240} [-20 \malpha_{6,1} - \malpha_{6,3} + \mC_{1}, \malpha_{6,2} + \mC_{2} ]
+ \mathcal{O}[(\Dxk)^{7}].
\end{multline}
Here,
\begin{multline}
\malpha_{6,1} = \Dxk \mjac_{2}, \qquad 
\malpha_{6,2} = \frac{\Dxk \sqrt{15}}{3} (\mjac_{3} - \mjac_{1}), \\
\malpha_{6,3} = \frac{10 \Dxk}{3} (\mjac_{3} - 2 \mjac_{2} + \mjac_{1}),
\end{multline}
while
\begin{equation}
\mC_{1} = [\malpha_{6,1}, \malpha_{6,2}], \qquad
\mC_{2} = -\frac{1}{60} [\malpha_{6,1}, 2\malpha_{6,3} + \mC_{1}],
\end{equation}
and $\mjac_{i} \equiv \mjac(\xk_{i})$ ($i=1,2,3$) are the Jacobian
matrices evaluated at the three Gauss-Legendre nodes within the
subinterval,
\begin{multline}
\xk_{1} = \xk + \left( \frac{1}{2} - \frac{\sqrt{15}}{10} \right) \Dxk, \qquad
\xk_{2} = \xk + \frac{\Dxk}{2}, \\
\xk_{3} = \xk + \left( \frac{1}{2} + \frac{\sqrt{15}}{10} \right) \Dxk.
\end{multline}

\label{lastpage}

\end{document}